\documentclass[aps,prb,superscriptaddress,shortbibliography,twocolumn]{revtex4-2}

\usepackage{lmodern}

\usepackage[latin9]{inputenc}
\usepackage{amsmath}
\usepackage{amssymb}
\usepackage{graphicx}
\usepackage{epstopdf}
\usepackage{color}
\usepackage{enumerate}
\usepackage{ulem}
\usepackage{eurosym}
\usepackage{fancyhdr}
\usepackage{layouts}
\usepackage{epstopdf}
\usepackage{textcomp}
\usepackage{anyfontsize}
\usepackage{setspace}
\usepackage{verbatim}
\usepackage[T1]{fontenc}
\usepackage{siunitx}
\usepackage[colorlinks=true]{hyperref} 
\usepackage{wrapfig}
\usepackage{caption}
\usepackage{ragged2e}

\usepackage{chemformula}

\begin{document}

\title{Growth dynamics of graphene buffer layer formation on ultra-smooth SiC(0001) surfaces}

\date{\today} 	

\author{Julia Guse}
\affiliation{Physikalisch-Technische Bundesanstalt, Bundesallee 100, 38116 Braunschweig, Germany}
\author{Stefan Wundrack}
\affiliation{Physikalisch-Technische Bundesanstalt, Bundesallee 100, 38116 Braunschweig, Germany}
\affiliation{LENA, Technische Universit\"at Braunschweig, Universit\"atsplatz 2, 38106 Braunschweig, Germany}
\author{Marius Eckert}
\affiliation{Physikalisch-Technische Bundesanstalt, Bundesallee 100, 38116 Braunschweig, Germany}
\affiliation{LENA, Technische Universit\"at Braunschweig, Universit\"atsplatz 2, 38106 Braunschweig, Germany}
\author{Peter Richter}
\affiliation{Institut f\"ur Physik, Technische Universit\"at Chemnitz, Reichenhainer Str. 70, 09126 Chemnitz, Germany}
\affiliation{Center for Materials Architectures and Integration of Nano Membranes (MAIN), Technische Universit\"at Chemnitz, Rosenbergstr. 6, 09126 Chemnitz, Germany}
\author{Susanne Wolff}
\affiliation{Institut f\"ur Physik, Technische Universit\"at Chemnitz, Reichenhainer Str. 70, 09126 Chemnitz, Germany}
\affiliation{Center for Materials Architectures and Integration of Nano Membranes (MAIN), Technische Universit\"at Chemnitz, Rosenbergstr. 6, 09126 Chemnitz, Germany}
\author{Niclas Tilgner}
\affiliation{Institut f\"ur Physik, Technische Universit\"at Chemnitz, Reichenhainer Str. 70, 09126 Chemnitz, Germany}
\affiliation{Center for Materials Architectures and Integration of Nano Membranes (MAIN), Technische Universit\"at Chemnitz, Rosenbergstr. 6, 09126 Chemnitz, Germany}
\author{Philip Sch\"adlich}
\affiliation{Institut f\"ur Physik, Technische Universit\"at Chemnitz, Reichenhainer Str. 70, 09126 Chemnitz, Germany}
\affiliation{Center for Materials Architectures and Integration of Nano Membranes (MAIN), Technische Universit\"at Chemnitz, Rosenbergstr. 6, 09126 Chemnitz, Germany}
\author{Markus Gruschwitz}
\affiliation{Institut f\"ur Physik, Technische Universit\"at Chemnitz, Reichenhainer Str. 70, 09126 Chemnitz, Germany}
\affiliation{Center for Materials Architectures and Integration of Nano Membranes (MAIN), Technische Universit\"at Chemnitz, Rosenbergstr. 6, 09126 Chemnitz, Germany}
\author{Kathrin K\"uster}
\affiliation{Max-Planck-Institut f\"ur Festk\"orperforschung, Heisenbergstr. 1, 70569 Stuttgart, Germany}
\author{Benno Harling}
\affiliation{IV. Physikalisches Institut, Georg-August-Universit\"at G\"ottingen, Friedrich-Hund-Platz 1, 37077 G\"ottingen, Germany}
\author{Martin Wenderoth}
\affiliation{IV. Physikalisches Institut, Georg-August-Universit\"at G\"ottingen, Friedrich-Hund-Platz 1, 37077 G\"ottingen, Germany}
\author{Christoph Tegenkamp}
\affiliation{Institut f\"ur Physik, Technische Universit\"at Chemnitz, Reichenhainer Str. 70, 09126 Chemnitz, Germany}
\affiliation{Center for Materials Architectures and Integration of Nano Membranes (MAIN), Technische Universit\"at Chemnitz, Rosenbergstr. 6, 09126 Chemnitz, Germany}
\author{Thomas Seyller}
\affiliation{Institut f\"ur Physik, Technische Universit\"at Chemnitz, Reichenhainer Str. 70, 09126 Chemnitz, Germany}
\affiliation{Center for Materials Architectures and Integration of Nano Membranes (MAIN), Technische Universit\"at Chemnitz, Rosenbergstr. 6, 09126 Chemnitz, Germany}
\author{Rainer Stosch}
\affiliation{Physikalisch-Technische Bundesanstalt, Bundesallee 100, 38116 Braunschweig, Germany}
\author{Klaus Pierz}
\affiliation{Physikalisch-Technische Bundesanstalt, Bundesallee 100, 38116 Braunschweig, Germany}
\author{Hans Werner Schumacher}
\affiliation{Physikalisch-Technische Bundesanstalt, Bundesallee 100, 38116 Braunschweig, Germany}
\author{Teresa Tschirner}
\affiliation{Physikalisch-Technische Bundesanstalt, Bundesallee 100, 38116 Braunschweig, Germany}

\begin{abstract}

In this study the growth process of epitaxial graphene on SiC was investigated systematically. The transition from the initial buffer layer growth to the formation of the first monolayer graphene domains was investigated by various techniques: atomic force microscopy, low energy electron diffraction, low energy electron microscopy, Raman spectroscopy, scanning tunneling spectroscopy and scanning electron microscopy. The data show that the buffer layer formation goes along with a simultaneous SiC decomposition which takes place as a rapid step retraction of one specific type of SiC bilayer in good agreement with the step retraction model. Once the buffer layer coverage is completed, the resulting characteristic regular repeating terrace and step height pattern of one and two SiC bilayers turned out to be very stable against further SiC decomposition. The following initial growth of monolayer graphene domains occurs, interestingly, only along the two bilayer high terrace edges. This behavior is explained by a preferential SiC decomposition at the higher step edges and it has some potential for spatial graphene growth control. The corresponding earlier graphene growth on one terrace type can explain the different scanning tunneling spectroscopy nanoscale resistivities on these terraces. 

\end{abstract}

\maketitle

\section{Introduction}
Monolayer carbon sheets arranged in a hexagonal lattice known as graphene have been extensively studied in the last decades due to their outstanding electronic and optical properties \cite{Geim2007}. Among a variety of fabrication techniques the atmospheric graphitization of SiC wafer surfaces is a promising route to obtain large-area graphene layers on substrates for technological compatible devices fabrication \cite{Berger2004, Emtsev2008, Virojanadara2008}.
The thermal decomposition of the SiC (0001) surface results in a carbon-rich surface reconstruction, the so-called buffer layer (BL) or zerolayer graphene, of which about one third of the C atoms are covalently bound to the substrate \cite{Emtsev2008, Riedl2010}. At higher annealing temperatures a new BL is formed at the interface and the former one turns into the freestanding graphene layer on the surface \cite{Hannon2011}. \\
A refined technique is the so-called polymer-assisted sublimation growth (PASG) method which results in the formation of ultra-smooth homogeneous monolayer graphene (MLG) with superior properties as negligible small resistance anisotropy and the absence of bilayer graphene (BLG) \cite{MomeniPakdehi2018a, Kruskopf2016}. This material is excellently suited for the fabrication of graphene-based quantum Hall resistance standards which allows the reproduction of half the value of the von Klitzing constant within a high accuracy of 2 ppb \cite{Yin2022}. 
The PASG method is remarkable because step bunching of the SiC substrate surface at the high growth temperature is suppressed by the rapid formation of the BL which stabilizes the SiC surface by its sp3 bonds. This is realized by the extra carbon supply from cracked polymer molecules alongside the silicon sublimation related carbon release from the SiC. Further, the growth method results in homogenous epitaxial MLG with atomic step heights of one and two SiC layers. 
\begin{minipage}{\linewidth}
      \centering
      \includegraphics[width=8cm]{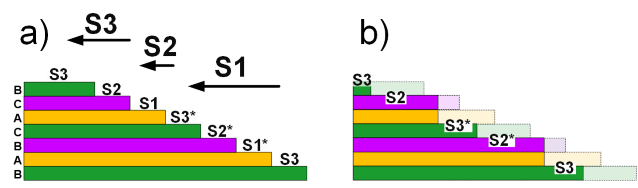}
      \captionof{figure}{Sketch of the step retraction model for SiC. \textbf{(a)} In the initial state the SiC surface consists of three equivalent terraces S1, S2 and S3 which are separated by SiC monolayer steps of about 0.25 nm. The arrow lengths indicate the different retraction velocities of the three terraces types during high temperature annealing. \textbf{(b)} At first the S1 terraces retract leaving behind a periodic pattern of S2 and S3 terraces with step heights of one and two SiC bilayers (about 0.5~nm and 0.25~nm). The simultaneously forming BL on the terraces is not shown.}
      \label{Figure0}
    \end{minipage}
The growth was explained by the step retraction model (see Fig.~\ref{Figure0} in which the three inequivalent SiC layers in the 6H-SiC unit cell, often called S1, S2 and S3, of the initial monoatomic stepped SiC surface retract with different velocity during the high temperature Si sublimation process \cite{Hupalo2009, Yazdi2013, MomeniPakdehi2020}. 
Under the well-controlled PASG growth conditions it was proposed that the typical terrace morphology of alternating step heights of one and two SiC layers is formed and frozen-in by the rapid forming BL \cite{Kruskopf2016, MomeniPakdehi2018a, MomeniPakdehi2020}. The existing growth models are not applicable, e.g, the step flow model examines the decomposition of three SiC bilayer high steps or even higher ones \cite{Ohta2010, Emtsev2009}. The diffusion controlled growth at shallow stepped SiC surfaces \cite{Emtsev2009} and the step-motion model \cite{Borovikov2009} describe the formation of typical "finger"-defects which is not observed here under optimized growth conditions \cite{Ohta2010}. \\
In this paper we investigate the early stage of PASG graphene growth. Firstly, a sample series of graphene on 6H-SiC with increasing growth duration was investigated by atomic force microscopy (AFM) giving insight into the temporal evolution of early BL growth. The data show that in the early growth stage, already before the buffer layer growth is completed, the typical terrace structure of alternating step heights of one and two SiC layers forms. Once the surface is completely covered with BL, the first MLG domains were observed which is verified by angle-resolved photoemission spectroscopy (ARPES), Raman spectroscopy, low-energy electron diffraction (LEED) and low-energy electron microscopy (LEEM) measurements. Detailed analysis of AFM, Scanning Electron Microscopy (SEM), and Scanning Tunneling Microscopy (STM) measurements revealed that formation of MLG domains predominantly occurs at higher steps corresponding to two SiC bilayers. This new and interesting behaviour is explained by the higher carbon supply from the decomposition of the corresponding terraces.  




\section{Sample preparation}

The graphene samples were fabricated by the PASG method on 6H-SiC substrates ($5 \times 10$~mm$^2$). The used epi-ready SiC wafer (II-VI Inc.) had a nominal miscut of $0.06^\circ$ against the [1-100] and about $0^\circ$ against the [11-20] crystal direction which guarantees an initial substrate surface of single SiC bilayer steps of 0.25~nm in height, and in direction close to a main crystallographic axis, verified by AFM inspection.  The substrates were cleaned in ultrasonic baths of acetone and isopropanol, 15 min each, and heated for 5~s at $800~^\circ$C. The diluted polymer (AZ5214E photoresist/isopropanol = 0.0034) was deposited by spin-on deposition on the Si-face of the SiC substrate, which allows for a high reproducibility of the same polymer coverage \cite{Chatterjee2022}. The subsequent graphene BL growth was performed in Ar atmosphere in an inductively heated reactor \cite{Ostler2010}. The temperature was raised at a rate of $\approx13$~K/s to the annealing temperature $T_a$ and stabilized for a time $t$. The growth parameters of the four samples in this study are listed in Table \ref{table:1}. 
\begin{table}[h!]
\setlength{\arrayrulewidth}{0.3mm}
\setlength{\tabcolsep}{6pt}
\renewcommand{\arraystretch}{1.2}
\begin{tabular}{|c|c|c|c|c|}
\hline
Sample &$T_a~(^\circ C)$ &$t~(min)$&$p_{Ar}~(mbar)$ \\
\hline
1 & 1400 &2 &900 \\
\hline
2 & 1450 &7 &900 \\
\hline
3 &1450 &10 &900 \\
\hline
4   &1400 &30 &500 \\
\hline
\end{tabular}
\caption{\justifying Growth parameters (temperature, growth time, and Ar partial pressure) of the four samples (shown in Fig.~\ref{Figure1}~(a-d)) investigated in this study.}
\label{table:1}
\end{table}

\vspace{5em}

\section{Results}

\begin{figure*}[t]
    \centering
	\includegraphics[width=15cm]{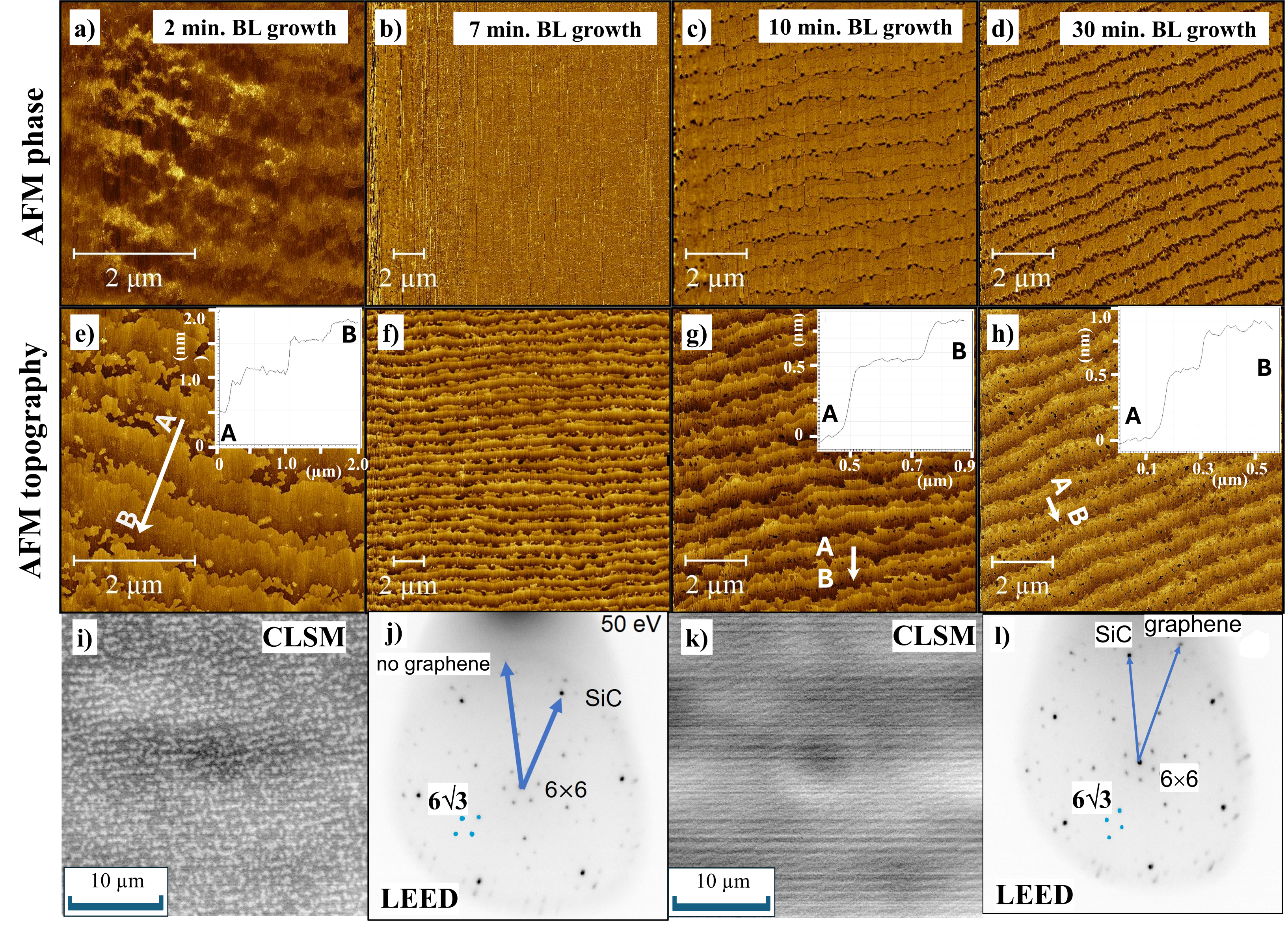}
        \captionof{figure}{\justifying \textbf{Buffer layer growth with varying parameters:} \textbf{(a-d)} AFM phase: BL grown with parameters of growth time $t$ = 2 min, 7 min, 10 min and 30 min, respectively, and different annealing temperatures $T_a$. \textbf{(e-h)} AFM topography: different step heights are revealed by line profiles highlighted in white and shown in the insets (A to B). \textbf{(i,k)} CLSM image: light spots represent elongated BL patches along the terraces (left panel) that grow together to form a uniform buffer layer with increasing growth time (right panel). \textbf{(j,l)} LEED: the pattern for 50 eV shows no graphene spot with 7 min growth time (left) and a graphene spot appearing with 30 min growth time (right).}
	\label{Figure1}
\end{figure*}

\indent The change of the surface morphology during BL growth was observed by AFM measurements of the four samples, see Fig.~\ref{Figure1}. Besides the height information from the topography images, differences in the surface material distribution (SiC, BL, MLG, BLG) can be obtained from the phase images. However, a clear assignment is not possible \cite{Lavini2020}. The initially equally stepped surface undergoes significant changes after applying the usual BL growth temperature of $1400~^\circ$C for 2~min \cite{Kruskopf2016}. Due to the short annealing time of 2 min no continuous BL has formed but elongated BL patches along the terraces are detected as bright spots in the confocal laser scanning microscopy (CLSM) image Fig.~\ref{Figure1}~(i) and as bright areas in the AFM phase image in Fig.~\ref{Figure1}~(a). There are no indications of residual polymer droplets showing that it has completely decomposed into C atoms supporting the graphitization \cite{Kruskopf2016}. The height profile of the AFM topography in Fig.~\ref{Figure1}~(e-h), reveals typical step heights of 0.25~nm and 0.5~nm corresponding to the height of one and two SiC bilayers, respectively. The appearance of double steps indicates that one SiC layer has been decomposed very early in the growth process. This layer with the highest retraction speed was attributed to the S1 terrace in the step retraction model.
The resulting surface topography suggests two types of terraces, narrower and wider, in a periodic sequence which were attributed to the S2 and S3 terraces located above the 0.5~nm and 0.25~nm steps, respectively \cite{MomeniPakdehi2020}.
%
%
It is interesting to see that this pattern is starting to form already at this early stage of BL growth. 
The AFM image (Fig.\ref{Figure1}~(e)) shows rough and irregular terrace edges and islands indicating the ongoing SiC decomposition at the terrace edges. Except the retraction of the S1 related step no further step bunching is observed which is attributed to the rapid formation of the buffer layer and the stabilization of the SiC (0001) surface by the covalent bonds to the SiC surface \cite{Kruskopf2016}. 
The formation of a homogenous BL takes places for slightly higher temperature and longer growth times. After 7~min growth time a nearly uniform phase image is observed, see Fig.~\ref{Figure1}~(b). Since only phase jumps at the terrace edges related to height jumps are detectable, a barely visible contrast is observed between neighboring terraces. The LEED pattern of this sample demonstrates the formation of the BL (see Fig.~\ref{Figure1}~(j)) with the typical ($6\sqrt{3} \times 6\sqrt{3}$)R30$^{\circ}$ reconstruction spots. The topography in Fig. \ref{Figure1}~(f) shows an slight improvement of the terrace uniformity. This indicates a continuing SiC decomposition and step retraction which has supplied additional carbon for the formation of the BL. This is in good agreement with another graphene growth study where it was shown that the graphene formation is supplied by carbon from two sources, the decomposition of the polymer as well as the SiC \cite{Chatterjee2022}. Since an ongoing SiC decomposition is observed, it is supposed that at this stage the polymer related carbon is completely consumed for the BL formation. With increasing growth time, after 10~min, the AFM topography in Fig.~\ref{Figure1}~(g) reveals a further improvement of the terrace uniformity. Compared to the early stages of growth, the periodical sequence of the two types of terraces is clearly observed now as well as the corresponding 0.25 and 0.5~nm high step edges.  Interestingly, an additional, new feature arises in the AFM phase image, Fig.~\ref{Figure1}~(c), which are isolated dark spots along the terrace edges. These spots are attributed to the formation of first MLG islands since the BL was already completed before. 
This is supported by the increasing size and density of the graphene islands on sample 4 with 30 min annealing time at $1400~^\circ$C. The characteristic graphene LEED spot pattern clearly proves that single layer graphene has formed on this sample to a certain extent \cite{Riedl2010}. The AFM phase image in Fig.~\ref{Figure1}~(d) shows a dark stripe consisting of these small islands along the higher terrace edges which will be discussed in detail later (Fig.~\ref{Figure4} and Fig.~\ref{Figure5}). Again the periodically repeated 0.25/0.5~nm terrace step pattern occurs in the AFM topography image in Fig.~\ref{Figure1}~(h). This clearly indicates that at this growth stage the carbon originating from the cracked polymer molecules by graphitization at high temperatures \cite{Kruskopf2016} is completely used up for the BL growth and the MLG is supplied by carbon from SiC decomposition.\\
\begin{figure*}[t]
	\includegraphics[width=15cm]{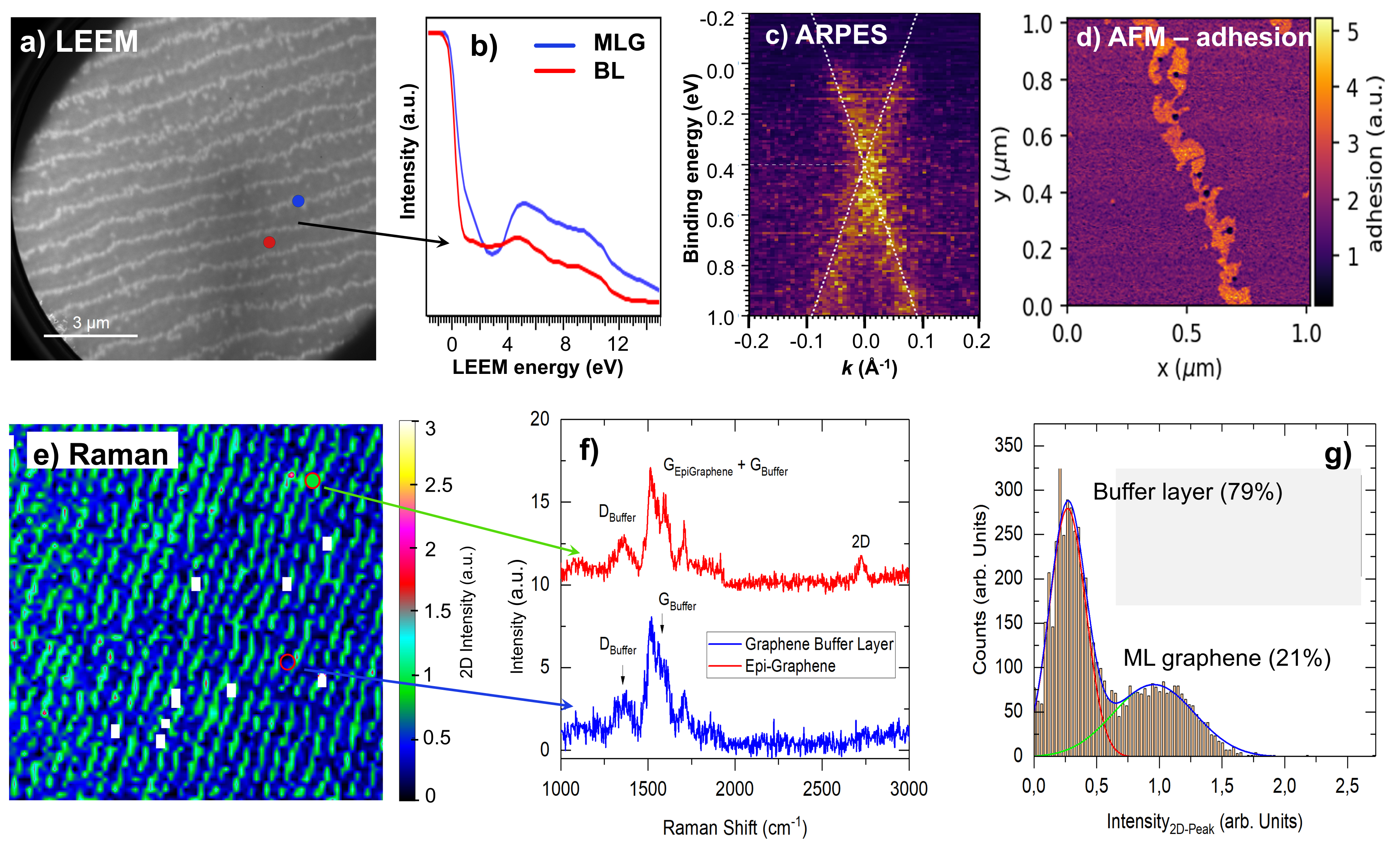}
		\caption{\justifying \textbf{Spectroscopic measurements:} \textbf{(a)} Bright field low-energy electron microscopy (BF-LEEM) image taken at 1.4 eV showing bright stripes (chain of graphene islands) on a  grey (BL) background. Scale bar: 3~$\mathrm{\mu m}$. \textbf{(b)} reflectivity spectra taken at two positions. Blue showing a BL signature and red a MLG signature. \textbf{(c)} Angle-resolved photemission spectroscopy (ARPES) measurement revealing a Dirac energy of 400~meV. \textbf{(d)} Peak-force tapping AFM measurements. The adhesion is used to discriminate between BL and MLG, showing the orange area as a MLG stripe on the terrace edge. \textbf{(e)} Raman mapping ($20 \times 20 ~\mathrm{\mu m}^2$) of the 2D intensity showing the coverage of BL and MLG. In \textbf{(f)}, two individual spectra taken from BL areas (blue) and MLG areas (red). \textbf{(g)} Extracted coverage of BL and MLG in \% from a histogram of the Raman measurement in \textbf{(e)}.}
	\label{Figure2}
\end{figure*}
Several experiments have been performed on this sample to investigate the nature of the dark spots, see Fig.~\ref{Figure2}. In the LEEM image the chain of graphene islands is discernible as nearly equidistant bright stripes comparable to the phase image in Fig.~\ref{Figure1}~(a). For a closer inspection the reflectivity spectra are taken at the position of the bright lines and at the terraces with darker contrast Fig.~\ref{Figure2}~(b). The pronounced dip in the spectra at about 3~eV indicates that the line of bright spots consists of MLG, while the terraces (no pronounced minimum in the I-V spectra) are covered by BL \cite{Ostler2014}. The absence of a second minimum also excludes the presence of BLG at the terrace edges. \\
The existence of MLG on the sample is also supported by the ARPES measurement at the K-point of the graphene Brillouin zone (Fig.~\ref{Figure2}~(c)). Here, the linear dispersing $\pi$-bands characteristic for graphene were found. The tight-binding approximation (white dashed lines as a guide to the eye) reveals a Dirac energy of 400~meV, which is in good agreement with earlier reports for MLG \cite{Riedl2009}. Thus, additional charge carriers associated with the local density of states (DOS) at the $(6\sqrt{3} \times 6\sqrt{3})$R30$^{\circ}$ interface may also contribute, potentially shifting the Dirac energy to 400~meV. \\
The nano-mechanical property of the surface is investigated by AFM in the material sensitive peak force tapping mode \cite{Ke2018}. Fig.~\ref{Figure2}~(d) displays the measured surface adhesion map from which the BL and MLG can be clearly resolved. The MLG domains show an irregular shape in high resolution, which line up along the terrace edge. The MLG patches in this growth state do not form a continuous graphene ribbon. \\
Additionally, Raman measurements were performed which allow a clear distinction between the different coverages with graphene. The observation of the 2D peak at about $2800~$cm$^{-1}$ with a full-width at half maximum (FWHM) smaller than $40~$cm$^{-1}$ is a characteristic signature of MLG underpinning the resonance conditions which are absent in BL \cite{Fromm2013}.  
\begin{figure*}[t]
	\includegraphics[width=13cm]{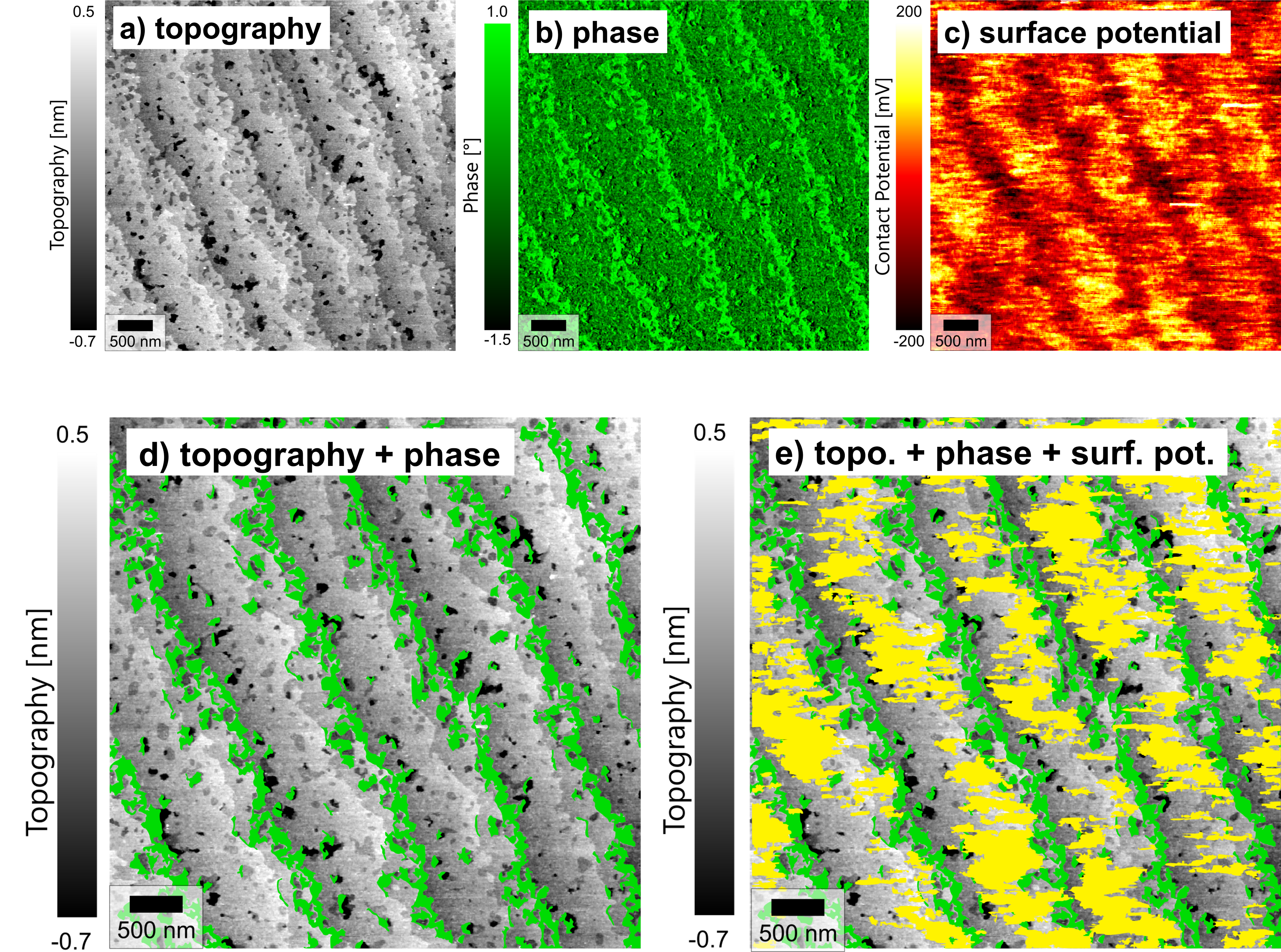}
		\caption{\justifying \textbf{AFM/KPFM measurements:} \textbf{(a)} AFM topography scan 10 $\times$ 10 $\mu m$. \textbf{(b)} AFM phase scan of the same position taken simultaneously. \textbf{(c)} KPFM measurement showing the surface potential. \textbf{(d)} Overlay of the topography scan and the phase scan to highlight the position of the graphene domains. \textbf{(e)} Overlay of topography, phase and surface potential.}
	\label{Figure3}
\end{figure*}
The Raman mapping in Fig.~\ref{Figure2}~(e) shows the intensity of the characteristic 2D peak evaluated from 10.000 Raman spectra on an area of $20 \times 20 ~\mathrm{\mu m}^2$. Here, the green areas indicate the MLG domains which show the elongated shape along the terrace edges. On the remaining surface (blue and black areas) no 2D peak is detected and there, the SiC is covered only with BL which gives rise to the broad BL related $D_{\mathrm{BL}}$ and $G_{\mathrm{BL}}$ band from 1350 to $1600~$cm$^{-1}$ related to local phonon DOS states \cite{Fromm2013}. The histogram of the 2D peak intensities from this surface area shows two maxima which were fitted by normal distributions. From the high intensity distribution which is related to MLG a coverage of 21~\% MLG is estimated. \\
The presented measurements clearly reveal that in the early growth stage when using the PASG method the SiC surface is homogenously covered with buffer layer graphene and initial MLG domains are forming at terrace steps. The exact locations of the graphene domains are investigated further by AFM, SEM and STM.  \\
The high-resolution AFM topography image in Fig.~\ref{Figure3}~(a) shows the two types of alternating terraces with the corresponding steps in height of one and two SiC bilayers in front of the wider and the narrower terraces, respectively. The combination of topography and phase image in Fig.~\ref{Figure3} reveals in which areas the graphene domains (light green areas) were formed. Interestingly, the domains were not formed to the same extent along both types of terrace steps. Most of the graphene was observed at the terrace with the higher step edge which is in front of the narrow terrace. Near the shallow step edge (adjacent to the wider terrace) much less and only smaller graphene islands are observed. In other words, only one in every two terrace steps have a significant share to the graphene formation. This is in contrast to other studies in which graphene formation was observed at each terrace \cite{Emtsev2009, Ohta2010}. But there, the terrace steps are 3 SiC bilayers in height or higher and a different growth mode is applicable. 
Moreover, holes with depths in the sub-nanometer range were found on the terraces where the SiC was decomposed locally. In the corresponding phase image, the contrast of the graphene domains is visible (marked in light green colour) located in the craters, preferably at the inner edges. Apparently, at higher step edges the SiC was decomposed preferably and graphene is formed \cite{Emtsev2009}.
The observed contrast in the KPFM image was additionally added to the AFM image in Fig.~\ref{Figure3}~(e). The positive potential (yellow contrast) was found on the BL terraces where the negative potential (dark red to black) is related to the graphene domains. The surface potential contrast confirms the distribution of MLG and BL. The difference between the contact potential acquired on the buffer layer and the already formed graphene ($+400~mV$) is in the order of previously reported values ($+600~mV$) \cite{Mammadov2017}. \\
\begin{figure*}[t]
	\includegraphics[width=15cm]{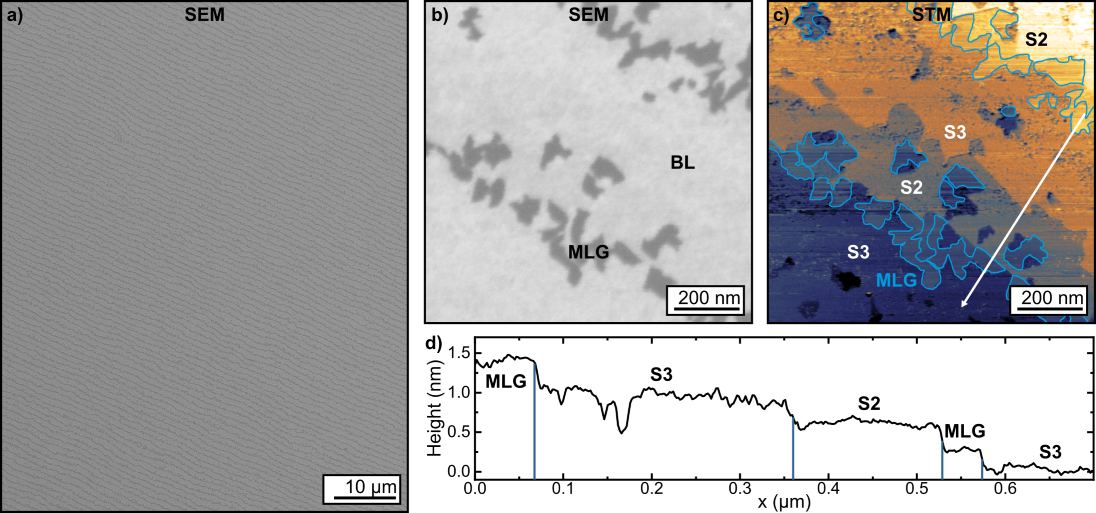}
		\caption{\justifying \textbf{SEM/STS measurements:} \textbf{(a)} Large scale SEM mapping shows uniform terraces covered by BL and small fractions of MLG. On the sub-$\mu m$ scale correlative SEM \textbf{(b)} and STM \textbf{(c)} reveals a more complex structure. The STM shows a superposition with the shape of the SEM MLG domains marked as blue lines. \textbf{(d)} The line profile along the marker in \textbf{(c)} reveals characteristic step heights between adjacent S2 and S3 terraces. MLG regions are only found on top of S3 terraces suggesting a faster decomposition rate. Both terrace types not only decompose from the edge inwards but also from within the terrace (most likely at defects). SEM images were measured at 15~kV, 1~nA. STM was measured at 4~V, 70~pA.}
	\label{Figure4}
\end{figure*}
The graphene distribution along the terrace steps is also observed in the SEM image in Fig.~\ref{Figure4}~(a) and (b). While the large scale SEM image Fig.~\ref{Figure4}~(a) underlines the excellent uniformity of the sample and a detailed scan in Fig.~\ref{Figure4}~(b) reveals MLG patches in good agreement with AFM and LEEM. The occurrence of isolated MLG domains (dark areas) is in agreement with the AFM and LEEM results. From the STM mapping in Fig.~\ref{Figure4}~(c) of the same area as measured by SEM shown in Fig.~\ref{Figure4}~(b) the heights of the corresponding structures are accessible. The STM image reveals a stronger roughness on the terraces potentially suggesting incomplete BL formation and three corrugated step edges between the corresponding terraces which are labelled S2 and S3 according to the step retraction model. Moreover, a strong decomposition of the S2 terraces is observed forming extended craters with MLG located at the inner edges. 
There is no indication of the three SiC terrace types since the S1 terraces with the fastest step retraction speed have already been decomposed in the very early stage of buffer layer growth as already deduced from the AFM measurements in Fig.~\ref{Figure1}. The remaining two terraces S2 and S3 are then related to 0.25~nm and 0.5~nm high steps. The location of the MLG domains can be clearly identified by comparison with the dark graphene areas in the SEM image (see light blue outlines in Fig.~\ref{Figure4}~(c)). It clearly shows that the MLG domains are located on the S3 terraces at the double step and not at the single step on the S2 terraces. This is confirmed by the detailed analysis of the cross-section profile in Fig.~\ref{Figure4}~(d). The blue lines indicate the step edges and it is apparent that only in front of the higher S2 terrace step the MLG domains are formed. In front of the lower 0.25~nm step almost no overgrown MLG areas were found suggesting that the probability for the decomposition of single stepped terraces is much lower or only enough carbon for BL formation was provided but not enough for MLG.
This is not clear at first sight since according to the step retraction model the S3 terrace should decompose faster and retract faster than the S2 terrace. However, this is valid only in the initial growth stage when comparing the individual S1, S2 and S3 terraces before step retraction has started. A less effective step retraction is also underlined by the formation of decomposition craters within S2 terraces rather than continuous decomposition at the step edges. In this intermediate growth state of two remaining terraces, it is assumed that the decomposition of the higher step edges is favourable since bond strength is reduced at the edges resulting in an increased surface energy and the decomposition of the S1 layer below S3 continues. The decomposed SiC double layer supplies sufficient carbon atoms to form the MLG domains at the step edges which is, however, not sufficient for the formation of a continuous graphene nanoribbon which would require the decomposition of 3 SiC bilayers \cite{Ohta2010, Yu2018}. 
The observed terrace morphology of periodically repeating S2 and S3 terraces with 0.25 and 0.5 nm high step edges obviously is very stable. It is still observed when the temperature is raised to more than $1600^{\circ}$C when homogenous monolayer graphene is grown \cite{MomeniPakdehi2020}. The described PASG method is a convenient way to prepare buffer layer or MLG on alternating S2 and S3 SiC terraces. Interestingly, some properties of EG on both surfaces are not identical, e.g., they show a difference in the nanoscale resistance observed by STS \cite{Sinterhauf2020}. This was related to the observed different structural modulations in the 10~pm-range of both SiC surfaces, namely a smoother topography of the S3 compared to the S2 terrace. As an origin for the potential modulation the partial Si depletion in the upper SiC substrate layer was supposed \cite{Gruschwitz2019}. With the presented study the reason for a different structural modulation of S2 and S3 can be found in the growth process. The faster growth of the graphene layer on the S3 terrace, observed in this study, acts as a capping layer protecting the underlying SiC for further decomposition since the probability of Si release is strongly reduced. Therefore, S3 should undergo a reduced Si depletion compared to S2 and should result in a smoother topography modulation as observed.

\section{Summary}

In this study epitaxial graphene on SiC was grown by the PASG method and we studied in detail the buffer layer and the initial MLG formation. It was shown that already before a complete buffer layer has formed step retraction of one SiC bilayer occurs in agreement with the step retraction model. The remaining regular repeating S2 and S3 terraces with the corresponding steps of 1 and 2 SiC bilayers (0.25 and 0.5~nm) prove to be very stable during the ongoing SiC decomposition and graphene growth. This SiC surface stability against step bunching is attributed to the rapid buffer layer formation by the additional carbon supply of the cracked polymer molecules. Interestingly, all measurements show that the initial formation of graphene domains does not occur along all terraces but preferentially along the higher step edge in front of the S2 terraces. This apparently contradicting result regarding the step retraction model is explained by the faster decomposition of the double steps compared to single ones. The resulting faster graphene growth on the S3 which protects the underlying SiC against ongoing Si loss can explain the smoother SiC surface of S3 terrace observed in another study. The observed step height related graphene formation bears potential to control the spatial growth of graphene and nano ribbons.

\section{Methods:}


\textbf{AFM}: The Atomic Force Microscopy images in Fig.~\ref{Figure1} were recorded in tapping mode with a Park NX 10 scanning force microscope. \\

\textbf{Peak-force tapping AFM} measurements were performed on a Bruker Dimension Icon in the ScanAsyst mode. A force distance curve is recorded at every point and the adhesion, i.e. the largest attractive force in the retrace curve is used to discriminate between BL and MLG. \\

\textbf{KPFM}: Kelvin Probe Force Microscopy measurements utilize the combination of a conventional AM-AFM (Agilent 5600SL) in intermitted contact mode at ambient conditions with the commercially available option to observe the surface's electric potential with the KPFM-mode as described by \cite{Melitz2011}.
A conducting Al/Pt tip and a second feedback loop with a lock-in amplifier, applying a bias voltage to the tip, canceling out the difference of the local electric potential of the sample and the tip. The value of the applied bias voltage is the negative electric potential difference also called contact potential difference (CPD) of the tip and sample.
Without calibration the absolute value cannot be estimated, as the tip's electric potential is dependent on the unknown apex configuration. The relative values within one measurement are valid as long as the tip does not change its configuration during scanning. \\

\textbf{CLSM}: In this study, we utilized an Olympus LEXT OLS5100 system for our Confocal Laser Scanning Microscopy (CLSM) measurements. The laser intensity image is created by combining a series of images in the Z direction to form a 2D intensity image.\\

\textbf{Raman}: Confocal Raman spectroscopic measurements have been carried out by using a Witec Alpha 300 RA equipped with a: grating of 300 grooves per mm, a Nd:YAG laser with an excitation wavelength of 488 nm (2.54~eV) and a focal length of 600~mm. Raman mapping were done across an area of $(20 \times 20)~\mu m^2$ with a step resolution of $0.2~\mu m$. The laser power has been kept below 2~mW. \\

\textbf{ARPES}: Band structure investigations were done using angle-resolved photoelectron spectroscopy (ARPES) using monochromatic HeI or HeII emission line (21.22, 40.82 eV) from a SPECS UVS 300 radiation source. The photoelectrons were detected using a 2D CCD detector equipped to a Specs Phoibos 150 analyzer. All measurements were conducted at room temperature. \\

\textbf{LEEM/LEED}: LEEM investigations were conducted with a SPECS FE-LEEM P90 instrument. Micro-LEED images were obtained by limiting the incoming electron beam to the desired region of interest ($ \approx 300~nm$) with the help of an aperture. From a series of LEEM images recorded as a function of electron energy, LEEM-IV data were extracted for selected regions of the surface. Conventional LEED measurements were performed with an ErLEED 150 system. \\

\textbf{SEM/STS}: Correlative SEM/STM were realized in a Omicron four probe STM/SEM system at a base pressure of $2\times10^{-10}$~mbar. STM measurements was performed at a high bias voltage of 4~V and low tunneling currents of 70~pA using an electro-chemical etched tungsten tip. A Zeiss Gemini SEM column placed above the STM stage allows direct correlation with a resolution of 4 nm achieved by an in-lens back-scattered electron detector at a beam energy of 15~kV and 1~nA beam current.

\section*{Acknowledgements:}

This work was supported by the Deutsche Forschungsgemeinschaft (DFG) FOR5242 research unit: project No. Ku4228/1-1, Pi385/3-1, Se1087/16-1, Te386/22-1  We-1889/14-1, and the DFG Germany's Excellence Strategy EXC-2123 QuantumFrontiers No.~390837967.


\bibliographystyle{unsrt}
\bibliography{BL_v0}{}

\end{document}